# Test of the FDTD Accuracy in the Analysis of the Scattering Resonances Associated with high-$Q$ Whispering-Gallery Modes of a Circular Cylinder

Artem V. Boriskin,[1,*] Svetlana V. Boriskina,[2] Anthony Rolland,[3]
Ronan Sauleau,[3] and Alexander I. Nosich[1]

[1] Institute of Radiophysics and Electronics NASU, Kharkiv, Ukraine
[2] Dept. of Electrical and Computer Engineering, Boston University, Boston, MA, USA
[3] Institut d'Electronique et de Télécommunications de Rennes, Université de Rennes 1,
Rennes cedex, France.
[*] Corresponding author: a_boriskin@yahoo.com,   Home page: www.ire.kharkov.ua/dep12/MOCA

The objective of the paper is the assessment of the accuracy of a conventional FDTD code in the computation of the near and far-field scattering characteristics of a circular dielectric cylinder. We excite the cylinder with an electric or magnetic line current and demonstrate the failure of the two-dimensional (2-D) FDTD algorithm to characterize accurately the emission rate and the field patterns near high-$Q$ whispering-gallery-mode resonances. This is proven by comparison with the exact series solutions. The computational errors in the emission rate are then studied at the resonances still detectable with FDTD, i.e. having $Q$-factors up to $10^3$.

## 1. Introduction

Spherical, toroidal, circular-cylindrical and some other open dielectric resonators are known to support so-called Whispering-Gallery-Mode (WGM) natural oscillations [1]. The famous features of WGMs are tight confinement of their inner field near the resonator boundary and "exponentially-high" radiation $Q$-factors. Circular-disk dielectric resonators are the key components in many advanced optical circuits including filters, couplers, laser cavities, etc. [1, 2]. This is due to their ability to support the WGM-like natural oscillations provided that the disk thickness is not too small and the radius is much larger than the wavelength. Among new configurations not mentioned in [1], there are thin disks embedded into a photonic-crystal matrix [3], coupled into a cyclic photonic molecule [4], and built into a coupled-resonator optical waveguide [5]. These coupled microresonators may provide additional advantages such as further lowering of the thresholds of lasing [6,7], improvement of emission directionality [3], and low-loss bending [8].

The available lithographic, epitaxial, and etching technologies enable controlled fabrication of thin microdisk resonators of given size and thickness. Still, due to fine nature of the WGM effects exploited in the optical and photonic devices, their accurate preliminary modeling may provide great savings in the cost and time of design.

Among the available numerical techniques used in computational photonics is the finite difference time-domain (FDTD) method [9,10]. Although FDTD is well known as a powerful and flexible tool, it has drawbacks also widely discussed in the literature, namely the numerical dispersion, staircase boundary approximation, and back-reflection from the borders of the computational window. These drawbacks become critical if physical boundaries of studied object do not coincide with the mesh and/or if high-$Q$ resonances are involved [11-13]. Both problems are present if a WGM resonator is considered.

It is known that a thin microdisk resonator can be studied in 2-D formulation with its bulk refractive index replaced by the effective index provided that the radius is considerably larger that the thickness and the wavelength [14]. This is because the effective index is taken as the normalized propagation constant of one of the guided modes on the infinite dielectric slab of the same thickness and bulk refractive index [15]. Therefore accurate 2-D modeling of WGMs is important task in computational photonics. Reduction of dimensionality significantly reduces the time and memory consumption for FDTD algorithms and thus enables one to choose denser meshes and smaller time steps. This may improve the accuracy of simulations within reasonable timing [16,17]. Additionally, various techniques have been proposed to improve the performance of FDTD solvers (see e.g. [9,18] and references therein). As a result, reasonable agreement between FDTD and experiment is often reported [17]. Assessment of the FDTD algorithms accuracy performed by comparison with numerical algorithms based on different techniques is also available in the literature, e.g. [11,13]. Nevertheless, the domain of trusted applicability for the FDTD solutions is still not completely clear because in most cases they have been compared with other approximate solutions to the Maxwell equations that have no built-in criteria of accuracy. Exception is the paper [11], where FDTD and method-of-moments (MoM) results were compared with Mie-type series data for the backward scattering cross-sections of the magneto-dielectric hollow circular cylinder illuminated by the normally incident plane electromagnetic wave. Conclusions of [11] were far from optimistic: in fact, both MoM and FDTD showed 100% and larger errors in the vicinity of sharp resonances that were associated with the modes that were still not truly WGM ones.

## 2. Outline of the Solution

To shed further light on this issue and specifically target WGMs, we study numerically the near and far-field characteristics of a single dielectric circular cylinder illuminated by a parallel line source (point source in 2-D). We compare the results computed by the conventional FDTD code and by the analytical series solution. This enables us to assess the accuracy of conventional FDTD code in the vicinities of high-$Q$ WGM resonances. We also compute the complex-valued natural frequencies of the WGMs to estimate the associated radiation $Q$-factors. This is done by means of an iterative gradient-type algorithm applied to the rigorous characteristic equation,

$$J_m(ka\sqrt{\varepsilon})H'_m(ka) - \sqrt{\varepsilon}H_m(ka)J'_m(ka\sqrt{\varepsilon}) = 0, \quad (1)$$

where $J_m(.)$ and $H_m(.)$ are the Bessel and Hankel functions, $m = 0,1,...$, the prime denotes the differentiation in argument, $k$ is the free-space wavenumber, $a$ is the cylinder radius, and $\varepsilon$ is its dielectric constant.

Our in-house 2-D FDTD algorithm has been developed based on the standard method proposed by K.S. Yee [19]. The Cartesian mesh has been used for the computational space surrounded with the Perfectly Matched Layers (PML) [20]. As a primary source, we use a line current modulated in time with a Gaussian pulse and set the parameters of PML that provide the normal back-reflection from the layer boundary below -50 dB for the whole considered frequency range. The time is discretized in accordance to the Courant stability criterion.

As reference data, we take the analytical solution built by using the separation of variables. This solution has the form of infinite series whose coefficients depend on the cylindrical functions. As they can be computed with machine precision, the accuracy of the series algorithm is controlled by the truncation order, $N$, provided that it exceeds $ka\sqrt{\varepsilon}$. We keep $N = ka\sqrt{\varepsilon} + 20$ thus the accuracy is at least $10^{-8}$.

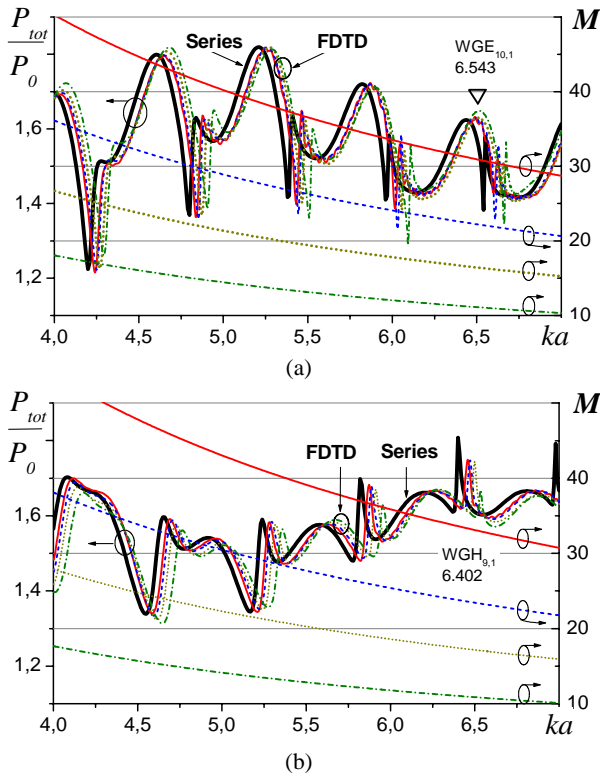

Fig. 1. Normalized total emission rate of the line source illuminating quartz ($\sqrt{\varepsilon} = 2.0$) circular resonator computed by FDTD method and Series (thick solid curves): (a) *E*–polarization and (b) *H*–polarization. The family of four FDTD curves is computed with different meshes. The values of *M* for each FDTD curve are represented by the monotonic lines of corresponding types.

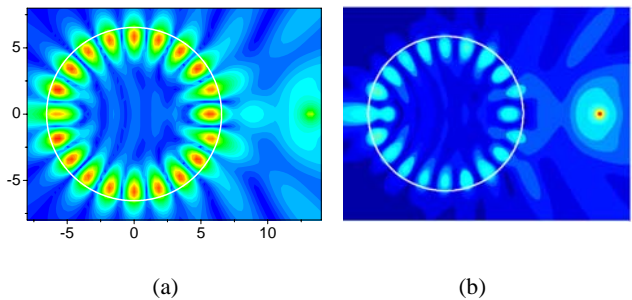

Fig. 2. Normalized near-field maps of quartz circular resonators excited by the *E*-polarized line source: (a) Series ($ka = 6.543$) and (b) FDTD ($ka = 6.600$). The corresponding WGM-$E_{10,1}$ resonance is indicated by the triangle in Fig. 1a.

## 3. Numerical Results

First, we consider a circular dielectric cylinder made of quartz ($\sqrt{\varepsilon} = 2.0$) excited by an electric or magnetic line current located at the distance equal to the resonator radius, $a$, from its boundary. Fig. 1 presents the total power radiated by the line source in the presence of cylinder normalized by the same value for a stand-alone line source,

$$P_0 = 2/Z_0 k \quad (E\text{-pol.}), \quad P_0 = 2Z_0/k \quad (H\text{-pol.}), \qquad (2)$$

versus the normalized frequency ($Z_0$ is the free space impedance). This quantity is the same as the normalized spontaneous emission rate in the source plus cavity system. In the analytical solution, it is reduced to the series with known coefficients. In the FDTD case, we computed the transient field values on auxiliary contour between the cylinder and the boundary of computational window and then integrated them into the total radiated power.

Results obtained with the FDTD and series algorithms relate to the left-side scale in Fig. 1, and the FDTD mesh size parameter, $M$, relates to the right-side scale. The thick solid line represents the reference series solution, while the family of four thin curves of different types is obtained by the FDTD algorithm with different mesh sizes taken as $\lambda_e/M$, where $\lambda_e$ is the wavelength in material. With a transient excitation, the FDTD method provides results over a wide frequency range in a single calculation. Although this is a great benefit comparing to the frequency domain methods, the accuracy of results differs within that range. It depends on the varying mesh size, which is indicated by four monotonically lowering curves of the corresponding types.

Periodic spikes, well seen in Fig. 1, correspond to the WGM-type $E_{m,1}$ and $H_{m,1}$ resonances. This is confirmed by their periodicity and the characteristic in-resonance near-field patterns presented in Fig. 2 for the $E$-case. In this notation, the indices indicate the number of the field function variations along the azimuth and the radius, respectively. As known, the higher the resonant frequency (equivalently, the denser the material and the larger the $m$), the higher the radiation quality factor $Q$. For the WGM resonances in the lossless cylinder ($\operatorname{Im}\sqrt{\varepsilon}= 0$), it behaves as $Q \sim \exp(\sqrt{\varepsilon} \operatorname{Re} ka)$, while for non-WGM ones it behaves as $Q \sim \sqrt{\varepsilon} \operatorname{Re} ka$.

Comparison of the curves obtained by two methods shows that the FDTD algorithm displays a regular shift of the whole curve to the higher frequencies. This can be explained by the staircase approximation of the cylinder boundary that is intrinsic for all conventional FDTD algorithms, so that actual domain filled in with dielectric is slightly smaller that the circle of radius $a$. Because of this shift, FDTD algorithm becomes especially inaccurate near the resonance frequencies. The shift can be reduced to a certain value by choosing a denser mesh. However, it cannot be eliminated completely and the minimum accessible error is apparently determined by the back-reflections from the virtual boundary of computational window; thus, it depends on the type of the non-reflecting boundary condition used and the shape and size of computational window. This conclusion is in line with earlier studies - e.g., see [11], [21].

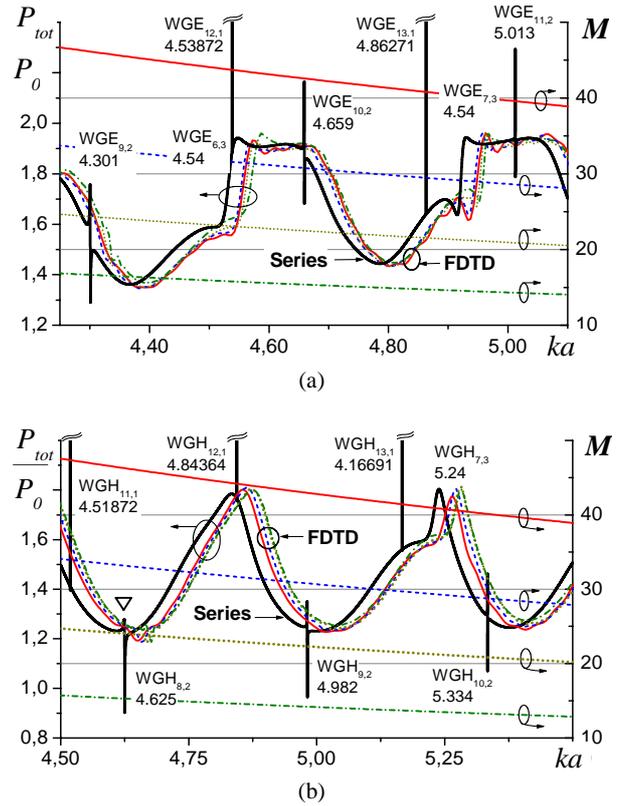

Fig. 3. The same as in Fig. 1 but for a silicon resonator ($\sqrt{\varepsilon} = 3.42$).

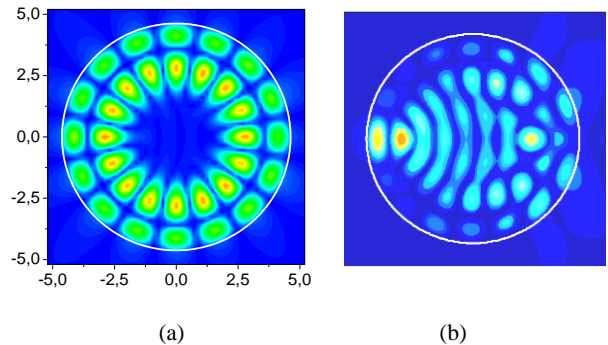

Fig. 4. Normalized near-field maps of the silicon circular resonators excited by a line $H$-polarized current: (a) Series ($ka$ = 4.625) and (b) FDTD ($ka$ = 4.645). The corresponding WGM-$H_{8,2}$ resonance is indicated by the triangle in Fig. 3b.

The curves in Fig. 3 are similar to Fig. 1 but computed for the line-source excited cylinder made of silicon ($\sqrt{\varepsilon} = 3.42$). Here, the radiation $Q$-factors of WGM-type natural modes are larger and accurate characterization of corresponding resonance spikes in the spontaneous emission rate with FDTD becomes troublesome. Note that for a silicon resonator higher-radial-order WGMs are also excited. Those detectable in Fig. 3 are of the first to the third order radial modes that is confirmed by the in-resonance near-field patterns presented in Fig. 4. Although the FDTD patterns in Figs. 2-b and 4-b have been computed at the corresponding (shifted) frequencies and with a dense mesh ($M = 50$), they are still quite inaccurate comparing to the exact ones computed by the series algorithm (Figs. 2-a and 4-a, respectively).

This reveals that, in principle, the spikes associated with higher-$Q$ resonances experience more difficulties when computed with FDTD algorithms. Therefore it is quite important to have *a priori* information about the quality factors of natural modes which can be involved in the solution of the considered problem in order to foresee the errors potentially entering the numerical solution. In general, this can only be done by solving the corresponding eigenvalue problem that may be difficult, especially for arbitrarily shaped resonators.

Here, note that FDTD algorithms are not able to solve the eigenvalue problems in direct manner – this has been emphasized in [2] in relation to the lasing modes. Instead, FDTD method always needs a pulsing source placed at some point inside or near a cavity and operates with the output in the form of time-domain signal computed at some other point (see, e.g., [22]). Further the $Q$-factors of detected natural modes are extracted from that signal by using Fourier transform. Therefore the result depends on the good or bad choice of the source and observation points and on the accuracy of numerically performed Fourier transform, to mention only the obvious factors. Such approach suffers of too many errors and may give only rough estimation of the associated $Q$-factors however still finds wide use in optical simulations.

The scattering analysis seems to be less amenable to such errors. However, as we will see below the matter is worse that it is normally assumed – FDTD apparently fails to detect the contribution of natural-mode spikes that have $Q$-factors larger than some threshold value.

In Figs. 5 and 6, we present the behavior of the FDTD computational errors in the frequency locations and in the absolute values, respectively, of the spikes of the normalized spontaneous emission rate corresponding to the WGM resonances visible in Fig. 1, i.e. for the quartz cylinder. They are presented as (discrete-valued) functions of the $Q$-factors of associated natural modes obtained by solving the equation (1). One can see that as soon as radiation $Q$-factor exceeds $10^2$, both errors start growing proportionally to the $Q$-factors.

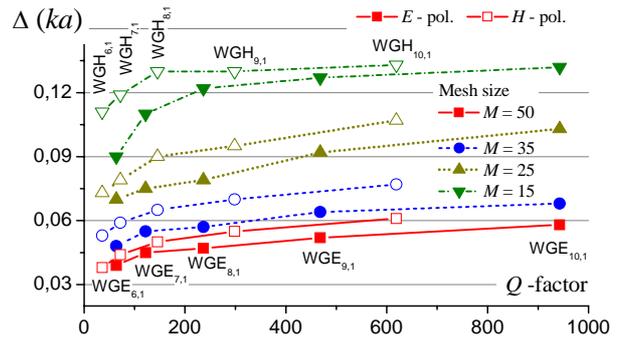

Fig. 5. The normalized frequency shift extracted from comparison between FTDT and Series solutions for the WGM-related spikes in the normalized emission rate observed in Fig. 1 for the quartz resonator vs. the resonances $Q$-factor. The family of four curves is for different mesh sizes.

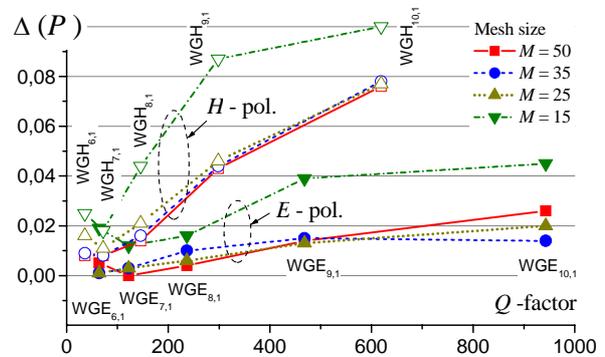

Fig. 6. Computation errors of the FDTD algorithm for the normalized emission rate extracted from comparison between FTDT and Series solutions presented in Fig. 1 for the quartz resonator vs. the resonances $Q$-factor. The family of four curves is for different mesh sizes.

**Table 1.** $Q$-factors of the ($m$,2) WGMs in a silicon circular resonator

| WGM-$E_{m,2}$ | | | WGM-$H_{m,2}$ | | |
|---|---|---|---|---|---|
| $m$ | $ka$ | $Q$ | $m$ | $ka$ | $Q$ |
| 9 | 4.30093 | 7862.90 | 8 | 4.62489 | 7380.38 |
| 10 | 4.65906 | 30946.36 | 9 | 4.98161 | 31393.24 |
| 11 | 5.01269 | 127414.73 | 10 | 5.33371 | 137516.08 |

Further, we have tried to extend the analysis of the FDTD errors to the resonances on the WGMs with higher $Q$-factors. As an example, exact radiation $Q$-factors of the modes having two field variations in radius, i.e. WGM-$E_{m,2}$ and WGM-$H_{m,2}$ with $m = 8$ to 11 are presented for the silicon cylinder in Table 1. We have found, however, that our FDTD algorithm is not capable of detecting the resonances on the WGMs with $Q$-factors higher than $10^3$. Therefore the higher-$Q$ resonances, for instance, those on the WGM-$E_{m,1}$ and WGM-$H_{m,1}$ modes that are observed in Fig. 3 on the curves computed via the series, are missing at all on the FDTD curves.

Note that radiation *Q*-factor values of the same range, i.e. several thousand, are characteristic not only to the WGM oscillations in the uniform circular or spherical cavities of moderate optical size but also to the WGMs in non-uniform discrete Luneburg lenses [23] and printed antennas on spherical substrates [24], bowtie modes in the stadium cavities [25], half-bowtie modes in the extended hemielliptic lenses [26,27], and many other important configurations. Therefore corresponding resonances in their scattering characteristics are not accessible with conventional FDTD codes and need finer simulation tools.

**Conclusion**

The accuracy of the FDTD algorithm as to the characterization of high-*Q* WGM resonances in the emission rate of a line source near a circular cylindrical resonator has been tested against the exact series solution. Comparison of numerical results confirmed reasonable accuracy of FDTD when applied out of high-*Q* WGM resonances, as well as a rapid growth of the computational error near to such resonances. Denser meshing reduces these errors to the level determined apparently by the type of absorbing boundary conditions used, shape and size of computational window, and other details of the FDTD code but does not eliminate them. The growth of the errors proportionally to the natural-mode *Q*-factors has been demonstrated. Finally, a complete failure of conventional FDTD code in the detection of WGM resonances with radiation *Q*-factors significantly higher than $10^3$ has been observed.